\documentclass[showpacs,a4paper]{revtex4}
\usepackage[english]{babel}
\usepackage[mathscr]{eucal}
\usepackage[dvips]{graphicx}
\usepackage{subfigure}
\usepackage[fleqn]{amsmath}
\def\ds{\displaystyle}
\def\bm#1{\hbox{\boldmath$#1$\unboldmath}}
\def\spacem#1{\setbox1\hbox{$#1$}\hbox to \wd1{\hfill}}

\def\hh{\hrule height0.9pt width1.1em}
\def\vv{\vrule width0.8pt depth0.12em height0.92em}
\def\square{\vbox{\kern0.15em\hh\kern0.9em\hh\kern-1.05em
            \hbox{\vv\kern0.93em\vv}}}
\def\frame#1#2#3#4{\vbox{\hrule height #1pt%
     \hbox{\vrule width #1pt\kern #2pt%
     \vbox{\kern #2pt%
     \hbox{\noindent#4}%
     \kern #2pt}%
     \kern #2pt\vrule width #1pt}%
     \hrule height0pt depth #1pt}}%

\begin{document}
\title{Spin and pseudospin symmetries of the Dirac equation with confining central potentials}
\author{P. Alberto}
\email{pedro.alberto@uc.pt}
\affiliation{Physics Department and Centro de F{\'\i}sica
Computacional, University of Coimbra, P-3004-516 Coimbra,
Portugal}
\author{A. S. de Castro}
\affiliation{Departamento de F{\'\i}sica e Qu{\'\i}mica,
Universidade Estadual Paulista, 12516-410 Guaratinguet\'a, S\~ao
Paulo, Brazil}

\author{M. Malheiro}
\affiliation{Departamento de F\'isica, Instituto Tecnol\'ogico de Aeron\'autica,
Centro T\'ecnico Aeroespacial, 12228-900 S\~ao Jos\'e dos Campos, S\~ao Paulo, Brazil}

\date{\today}

\begin{abstract}
We derive the node structure of the radial functions which are solutions of the Dirac equation with scalar $S$
and vector $V$ confining central potentials, in the conditions of
exact spin or pseudospin symmetry, i.e., when one has $V=\pm S+C$, where $C$ is a constant.
We show that the node structure for exact spin symmetry is the same as the one for central potentials
which go to zero at infinity but for exact pseudospin symmetry the structure is reversed.
We obtain the important result that it is possible to have positive energy bound solutions
in exact pseudospin symmetry conditions for confining potentials of any shape, including naturally
those used in hadron physics, from nuclear to quark models.
Since this does not happen for potentials
going to zero at large distances, used in nuclear relativistic mean-field potentials or in the atomic nucleus,
this shows the decisive importance of the asymptotic
behavior of the scalar and vector central potentials on the onset of pseudospin symmetry and
on the node structure of the radial functions.
Finally, we show that these
results are still valid for negative energy bound solutions for anti-fermions.
\end{abstract}

\maketitle

\section{Introduction}

Spin and pseudospin symmetries are $SU(2)$ symmetries of a Dirac Hamiltonian with vector and scalar potentials.
They are realized when the difference, $\Delta=V-S$, or the sum, $\Sigma=V+S$, are constants.
The near realization of these symmetries may explain degeneracies in some heavy meson spectra (spin symmetry) or
in single-particle energy levels in nuclei (pseudospin symmetry), when these physical systems are described by relativistic
mean-field theories (RMF) with scalar and vector potentials \cite{gino_prl,gino_rev_2005}.
Recently it was found that nuclear resonant states exhibit similar features as bound states, namely
that in conditions of pseudospin symmetry the same pseudospin quantum numbers will be conserved and the pseudospin doublets would
have the same energy and width \cite{Lu_Zhao_Zhou}. The spin and pseudospin symmetries for a Dirac
equation with central Coulomb potentials, together with the node structure of its radial function solutions
were also recently discussed in ref.~\cite{pedro_pra}.
When these symmetries are realized, they decouple the upper and lower
components of Dirac equation so its solutions behave, as far as the energy spectrum is concerned, as spin zero
solutions of the Klein-Gordon equation with the same vector and scalar potentials \cite{prc_75_047303}.
For systems whose potentials go to zero at infinity, pseudospin symmetry cannot be realized for positive energy
solutions but only for negative energy solutions \cite{Zhou,pedro_pra,He_Zhou_Meng,liang,ronai_wscc}.
The reverse is true for spin symmetry.
However, for harmonic oscillator potentials, also used as nuclear mean-fields,
one is able to find bound solutions
when either spin or pseudopin symmetries are realized \cite{Chen_Meng,harm_osc_prc_2004,prc_73_054309}.
In this paper we will derive the node structure of radial functions for central vector and scalar potentials which are finite at the
origin and go to infinity as $r\to\infty$ when these potentials satisfy spin and pseudospin conditions, independently of their shape.
We show in a very general way
that for these confining potentials it is possible to have positive energy bound solutions for exact pseudospin symmetry,
contrary to what happens for potentials going to zero at large distances, as is the case of nuclear RMF.
This quite general finding means that for these potentials it is possible to realize exactly this symmetry in nature,
which can be relevant also
in particle physics where confining potentials like the Cornell potentials are of great interest.
Finally, we also show that these results are still valid for bound states of anti-fermions, i.e.,
that those states can exist in the exact spin symmetry conditions.

\section{Node structure of the radial functions for confining potentials}
\label{sec:radial_nodes_dirac}
With no significant loss of generality, we will set $\Delta=0$ for exact spin symmetry and $\Sigma=0$ for exact
spin symmetry. The derivations in this section follow closely the procedure of Leviatan and Ginocchio
\cite{levi_gino}.

The Dirac Hamiltonian with scalar $S$ and vector $V$ potentials reads
\begin{equation}\label{H}
H=\bm\alpha\cdot \bm p\,c + \beta (mc^2 + S) + V\ ,
\end{equation}
where $\bm\alpha$ and $\beta $ are the Dirac matrices in the usual representation
\begin{equation}
\bm\alpha=\left(
\begin{array}{cc}
0 &\bm\sigma \\[1mm]
\bm\sigma & 0
\end{array}
\right), \qquad \beta = \left(
\begin{array}{cc}
I & 0 \\[1mm]
0 & -I
\end{array}
\right)\, ,
\end{equation}
where $\bm\sigma$ are the Pauli matrices and $I$ is the $2\times 2$ unit matrix.
The Hamiltonian (\ref{H}) can be written in terms of the sum and difference potentials $\Sigma=V+S$ and $\Delta=V-S$ as
\begin{equation}\label{H2}
H=\bm\alpha\cdot \bm p\,c + \beta mc^2 + \frac12(I+\beta)\Sigma + \frac12(I-\beta)\Delta\ ,
\end{equation}
The general solution of the time-independent Dirac equation
$H \psi = E \psi$ for central potentials is
\begin{equation}
\psi = \left(\begin{array}{c}
\ds i\, \frac{g_{\kappa}(r)}r\, \phi_{\kappa m_j}(\theta,\varphi)\\[3mm]
 \ds-\frac{f_{\tilde\kappa}(r)}r\,\phi_{\tilde\kappa  m_j}(\theta,\varphi)
 \end{array}\right)
\label{psi}
\end{equation}
where 
\begin{equation}
\label{def_kappa}
\kappa=\left\{
\begin{array}{cl}
- (\ell+1) & \quad j =  \ell + \frac12 \\[2mm]
   \ell  & \quad j  =  \ell - \frac12
       \end{array}\right.\ ,
\end{equation}
$\ell$ is the upper component orbital angular momentum and $\tilde\kappa=-\kappa$.
The angular functions $\phi_{\kappa m_j}(\theta,\varphi)$ are the spinor spherical harmonics
and $g_{\kappa}(r)$, $f_{\tilde\kappa}(r)$ are the radial wave functions for
the upper and lower components of the Dirac spinor respectively.
The orbital and total angular momenta can be obtained from $\kappa$ by $\ell=|\kappa|+1/2\big(\kappa/|\kappa|-1\big)$
and $j=|\kappa|-1/2\ .$
The radial functions $g_{\kappa}(r)$ and $f_{\tilde\kappa}(r)$ satisfy the coupled first-order  differential
equations
\begin{subequations}
\label{eq_radiais_1}
\begin{align}
\frac{d g_\kappa}{dr} + \kappa\, \frac{g_\kappa}{r} & =  A(r)\,f_{\tilde\kappa}\\
\frac{d f_{\tilde\kappa}}{dr} - \kappa\, \frac{f_{\tilde\kappa}}{r} & =  -B(r)\,g_\kappa \ ,
\end{align}
\end{subequations}
where
\begin{subequations}
\label{def_A_B}
\begin{align}
A(r) &= \frac1{\hbar c}\,[E+mc^2-\Delta(r)] \\
B(r) &= \frac1{\hbar c}\,[E-mc^2-\Sigma(r)]\ .
\end{align}
\end{subequations}
If the potentials $\Delta(r)$ and $\Sigma(r)$ are finite at $r=0$, then near the origin
one must have $g_{\kappa}(r)\propto r^\alpha$ and $f_{\tilde\kappa}(r)\propto r^\beta$ where $\alpha$ and $\beta$
are integers greater than or equal to 1. From eqs.~(\ref{eq_radiais_1}) one finds, when $r\to 0$,
\begin{subequations}
\label{g_f_origin}
\begin{align}
\left\{
\begin{array}{l}
g_{\kappa}(r)\propto r^{\kappa+1} \\[1.5mm]
f_{\tilde\kappa}(r)\propto r^{\kappa}
\end{array}\right.  & \qquad\kappa>0 \\[2mm]
\left\{
\begin{array}{l}
g_{\kappa}(r)\propto r^{-\kappa} \\[1.5mm]
f_{\tilde\kappa}(r)\propto r^{1-\kappa}
\end{array}\right.  & \qquad\kappa<0\ .
\end{align}
\end{subequations}
Setting $\mathcal{G}_\kappa=r^\kappa g_\kappa$ and $\mathcal{F}_{\tilde\kappa}=r^{\tilde\kappa} f_{\tilde\kappa}=
r^{-\kappa} f_{\tilde\kappa}$ eqs.~(\ref{eq_radiais_1}) can be written as
\begin{subequations}
\label{eq_radiais_2}
\begin{align}
\frac{d \mathcal{G}_\kappa}{dr} & =  r^{2\kappa}A(r)\,\mathcal{F}_{\tilde\kappa}\\
\frac{d \mathcal{F}_{\tilde\kappa}}{dr} & =  -r^{-2\kappa}B(r)\,\mathcal{G}_\kappa
\end{align}
\end{subequations}
The functions $\mathcal{G}_\kappa$ and $\mathcal{F}_{\tilde\kappa}$ have the same nodes as
$g_{\kappa}$ and $f_{\tilde\kappa}$ for $r>0$. As remarked in ref.~\cite{levi_gino},
eqs.~(\ref{eq_radiais_2}) imply that the nodes of $g_{\kappa}$ and $f_{\tilde\kappa}$
alternate, i.e., one function radial node is located between two
consecutive nodes of the other function.
One the other hand, if $r_1$ and $r_2$ are the the nodes of
$\mathcal{F}_{\tilde\kappa}$ and $\mathcal{G}_\kappa$ respectively,
one has
\begin{subequations}
\label{2nd_deriv_nodes}
\begin{alignat}{2}
\left.\frac{d^2 \mathcal{G}_\kappa}{dr^2}\right|_{r=r_1} & =
-A(r_1)B(r_1)\,\mathcal{G}_\kappa(r_1)&\quad\mathcal{F}_{\tilde\kappa}(r_1)&=0\\
\left.\frac{d^2 \mathcal{F}_{\tilde\kappa}}{dr^2}\right|_{r=r_2} & =
-A(r_2)B(r_2)\,\mathcal{F}_{\tilde\kappa}(r_2)&\quad\mathcal{G}_\kappa(r_2)&=0
\end{alignat}
\end{subequations}
From (\ref{eq_radiais_2}) and (\ref{2nd_deriv_nodes}) $\mathcal{G}_\kappa$ and $\mathcal{F}_{\tilde\kappa}$
have an extremum when the other function has a node. Moreover, since we will be looking for
bound solutions,
$\mathcal{G}_\kappa$ and $\mathcal{F}_{\tilde\kappa}$ must go to zero when $r\to\infty$
and so these extrema must be a maximum when the function is positive and a minimum when the function is
negative. Then, from (\ref{2nd_deriv_nodes}), one must have $A(r)B(r)>0$ at each node.
When there is spin symmetry ($\Delta=0$), $A(r)>0$, so that condition means that one must have
$B(r)=E-mc^2-\Sigma>0$  at the nodes of $\mathcal{G}_\kappa$ and $\mathcal{F}_{\tilde\kappa}$ which is to say that the kinetic
energy is positive, since $\Sigma$ acts as a binding potential.
This means that in this case all nodes occur within the classically allowed region, a situation
similar to the case when binding potentials go to zero, as remarked in ref.~\cite{levi_gino}.
In pseudospin symmetry conditions ($\Sigma=0$), and assuming that the confining potentials are
positive, one has $E> mc^2$ and therefore $B(r)>0$, so that $A(r)$ also has to be positive at the nodes of
 $\mathcal{G}_\kappa$ and $\mathcal{F}_{\tilde\kappa}$ as a necessary condition for the radial functions to go to zero at infinity.
Although $\Delta$ cannot be considered strictly a binding potential,
$A(r)$ can be related to a position dependent effective mass (see, for instance,
\cite{prc_65_034307}) and so the condition $E> -mc^2+\Delta$ or that $A(r)$ must be positive
at the nodes is equivalent to the ''classical" condition that the effective mass must
be positive at those nodes.

We turn now to the asymptotic behavior of radial functions considering that
the potentials $\Delta$ or $\Sigma$ are confining potentials, i.e., go
to infinity as $r\to\infty$.
To have bound solutions, the radial functions must go to zero
at infinity and,
due to the symmetry of eqs.~(\ref{eq_radiais_1}) we expect that both $g_{\kappa}(r)$ and $f_{\tilde\kappa}(r)$ will have similar
asymptotic behavior so that, when $r\to\infty$, those equations may be written as
\begin{subequations}
\label{eq_radiais_1_asymp}
\begin{align}
\frac{d g_\kappa}{dr} & =  A(r)\,f_{\tilde\kappa}\\
\frac{d f_{\tilde\kappa}}{dr} & =  -B(r)\,g_\kappa
\end{align}
\end{subequations}
and the corresponding 2nd-order equations as
\begin{subequations}
\label{eq_radiais_2_asymp}
\begin{align}
\frac{d^2 g_\kappa}{dr^2} & =  \frac1{A(r)}\frac{d A(r)}{dr}\,\frac{d g_\kappa}{dr}-A(r)B(r)\,g_\kappa\sim -A(r)B(r)\,g_\kappa\\
\frac{d^2 f_{\tilde\kappa}}{dr^2} & =  \frac1{B(r)}\frac{d B(r)}{dr}\,\frac{d f_{\tilde\kappa}}{dr}-A(r)B(r)\,f_{\tilde\kappa}\sim -A(r)B(r)\,f_{\tilde\kappa}
\end{align}
\end{subequations}
since the derivative term will go faster to zero than the function term. One sees immediately that
the product $A(r)B(r)$ must be negative asymptotically in order that the radial functions go to
zero at infinity because their second derivative must has the same sign as the function.

One may assume the following asymptotic form for the radial functions ($\lambda$ is a positive constant)
\begin{equation}\label{g_f_asymp}
g_{\kappa}(r) \sim f_{\tilde\kappa}(r) \sim e^{-\lambda f(r)}
\end{equation}
where $f(r)$ is a increasing function of $r$. Inserting this \textit{ansatz}
into eqs.~(\ref{eq_radiais_2_asymp}) we can determine $f(r)$ and the sign of the potentials. For instance, if
\begin{align}
\Delta(r)&\xrightarrow[r\to\infty]{} Cr^a\\
\Sigma(r)&\xrightarrow[r\to\infty]{}Dr^a
\end{align}
with $a> 0$ and $C$ and $D$ are constants, one would have $f(r)=r^{a+1}$ if both $C$ and $D$ are different from zero or
$f(r)=r^{a/2+1}$ if either $C$ or $D$ is zero, as would be the case for spin or pseudospin symmetry conditions, respectively.
For instance, if $a=2$ (harmonic oscillator potentials), in spin or pseudospin symmetry conditions, the
radial functions would behave as Gaussians, as indeed is the case (see \cite{harm_osc_prc_2004}).
As mentioned before, at the same time we get the sign of the potentials from eqs.~(\ref{eq_radiais_2_asymp}),
so that, again in spin or pseudospin symmetry conditions, one has respectively
either $D>0$  (from (\ref{def_A_B}) $B(r)$ would be a positive constant, because $E>mc^2$)
or $C>0$ ($A(r)$ would be a positive constant). This is, of course,
what one would expect for confining potentials for positive energy solutions,
either acting as binding potentials like $\Sigma$ or as an effective
mass like $\Delta$.

To obtain a relation between the nodes of $g_\kappa$ and $f_{\tilde\kappa}$ it is useful to consider the behavior
of the product $g_\kappa f_{\tilde\kappa}$. From eqs.~(\ref{eq_radiais_1}) one obtains
\begin{equation}
\label{deriv_gf}
\frac{d(g_\kappa f_{\tilde\kappa})}{dr}=A(r)f^2_{\tilde\kappa}-B(r) g^2_\kappa\ .
\end{equation}

Let us consider separately the cases for exact spin and pseudospin symmetries.

\subsection{Spin symmetry}

At $r\to 0$, the behavior of the radial functions (\ref{g_f_origin}) and eqs.~(\ref{def_A_B})
imply that
\begin{subequations}
\label{gf_origin}
\begin{align}
\frac{d(g_\kappa f_{\tilde\kappa})}{dr}&\sim -B(0)g^2_\kappa<0   \qquad\kappa<0\\
\frac{d(g_\kappa f_{\tilde\kappa})}{dr}&\sim A(0)f^2_{\tilde\kappa}>0 \qquad\kappa>0\ .
\end{align}
\end{subequations}
On the other hand, when $r\to\infty$, because $A(r)$ is constant for exact spin symmetry,
we get
\begin{equation}
\label{gf_inf_spin}
\frac{d(g_\kappa f_{\tilde\kappa})}{dr}\sim -B(r) g^2_\kappa > 0\ .
\end{equation}
Since $g_\kappa f_{\tilde\kappa}$ is zero at $r=0$, eqs.~(\ref{gf_origin}) give its sign near the origin
and the same happens with eq.~(\ref{gf_inf_spin}) at large $r$.
We have
\begin{subequations}
\label{gf_spin}
\begin{alignat}{2}
r&\to 0&\qquad g_\kappa& f_{\tilde\kappa}<0 \qquad(\kappa<0)\\
r&\to 0&\qquad g_\kappa& f_{\tilde\kappa}>0 \qquad(\kappa>0)\\
r&\to\infty&\qquad g_\kappa& f_{\tilde\kappa}<0
\end{alignat}
\end{subequations}
Moreover, from (\ref{deriv_gf}) one has at the nodes $r_1>0$ and $r_2>0$ of
$f_{\tilde\kappa}$ and $g_\kappa$ respectively
\begin{subequations}
\label{deriv_gf_nodes}
\begin{align}
\left.\frac{d(g_\kappa f_{\tilde\kappa})}{dr}\right|_{r=r_1} & =
-B(r_1) g^2_\kappa <0\\
\left.\frac{d(g_\kappa f_{\tilde\kappa})}{dr}\right|_{r=r_2} & =
A(r_2)f^2_{\tilde\kappa} >0\ ,
\end{align}
\end{subequations}
since, as referred previously, both $A(r)$ and $B(r)$ are positive at the nodes of the radial functions.

From (\ref{gf_spin}) we see that $g_\kappa f_{\tilde\kappa}$ changes sign when going from zero to infinity if $\kappa>0$
while it is negative near the origin as well as at large $r$ if $\kappa<0$. This means that $g_\kappa f_{\tilde\kappa}$
has an odd number of nodes (not counting the origin) when $\kappa>0$ and an even number of nodes
when $\kappa<0$. On the other hand, from (\ref{gf_inf_spin}) and (\ref{deriv_gf_nodes}) we see
that the last node is a node of $f_{\tilde\kappa}$ because the derivative at the last node must have the opposite sign
of the asymptotic derivative. By the same argument, from (\ref{gf_origin}) and (\ref{deriv_gf_nodes}),
when $\kappa>0$ the first node ($r>0$) must be a $f_{\tilde\kappa}$ node,
while when $\kappa<0$ the first node is a $g_\kappa$ node. Therefore, and since the nodes of $f_{\tilde\kappa}$
and $g_\kappa$ alternate, when $\kappa>0$ $f_{\tilde\kappa}$ must have one more node than $g_\kappa$ and when $\kappa<0$ the
radial functions have the same number of nodes. In summary, denoting by $n_f$ and $n_g$ the number of nodes of
$f_{\tilde\kappa}$ and $g_\kappa$ respectively, we have
\begin{align}
\begin{array}{ll}
n_f=n_{g} &\qquad\kappa<0\\[0.2cm]
n_f=n_{g}+1&\qquad\kappa>0\ .
\end{array}
\label{nf_spin}
\end{align}
This relation is known to hold also for central potentials that go to zero at
infinity as the nuclear mean-field potentials both for fermions and anti-fermions
\cite{gino_rev_2005,levi_gino,ronai_wscc}.

\subsection{Pseudospin symmetry}

In pseudospin conditions, the behavior of  $g_\kappa f_{\tilde\kappa}$ at the origin is
the same as for spin symmetry, but when $r\to\infty$ we have
\begin{equation}
\label{gf_inf_pspin}
\frac{d(g_\kappa f_{\tilde\kappa})}{dr}\sim A(r) f^2_\kappa < 0
\end{equation}
so that now we have
\begin{subequations}
\label{gf_pspin}
\begin{alignat}{2}
r&\to 0&\qquad g_\kappa& f_{\tilde\kappa}<0 \qquad(\kappa<0)\\
r&\to 0&\qquad g_\kappa& f_{\tilde\kappa}>0 \qquad(\kappa>0)\\
r&\to\infty&\qquad g_\kappa& f_{\tilde\kappa}>0
\end{alignat}
\end{subequations}

Equations (\ref{deriv_gf_nodes}) still hold in this case, so using the same reasoning as in the spin symmetry
case, we find that now the last node must be a $g_\kappa$ node, while the first node will still be
a $f_{\tilde\kappa}$ node if $\kappa>0$ and a $g_\kappa$ node if $\kappa<0$. So in this case we will have
\begin{align}
\begin{array}{ll}
n_f=n_{g}-1 &\qquad\kappa<0\\[0.2cm]
n_f=n_{g}&\qquad\kappa>0\ .
\end{array}
\label{nf_pspin}
\end{align}

In ref.~\cite{harm_osc_prc_2004} was shown that indeed one has such radial node structure for the harmonic oscillator
potentials $\Sigma$ and $\Delta$ in spin and pseudospin symmetry conditions. This structure can be understood by noting
that in pseudospin symmetry conditions the radial functions interchange roles relative to spin symmetry conditions,
so we have $g_\kappa\leftrightarrow f_{\tilde\kappa}$ and $\kappa\leftrightarrow \tilde\kappa=-\kappa$.

\section{Existence of bound solutions for exact pseudospin symmetry}

A well-known fact about pseudospin symmetry is that it cannot be exact in relativistic mean-field quantum systems
because a $\Sigma$ potential going to zero at infinity cannot not be zero everywhere to provide some binding, so that
one has $E<mc^2$. As remarked by Leviatan and Ginocchio \cite{levi_gino} the previous statement can be
rephrased by saying that in order to have bound solutions of the Dirac equation with scalar and vector
central potentials of that kind, the derivative of $g_\kappa f_{\tilde\kappa}$ at large $r$ must change sign
when $r$ decreases to be able to have a node or at least to go to zero at $r=0$. For nuclear mean-field
potentials this means that, from eq.~(\ref{deriv_gf}) and since $A(r)>0$ everywhere \cite{levi_gino},
$B(r)$ must be able to change its sign from the negative asymptotic value, i.e., one must have $B(r)>0$ for some $r$ values,
or, as noted before, there should be some region in which the kinetic energy is positive,
that is, a classically allowed region.

However, if one has confining potentials, one has at hand another mechanism to bind particles, namely
by letting the effective mass go to infinity as $r\to\infty$, which, as referred before, is equivalent
to have a confining $\Delta$ potential. In this case, one is allowed to let $\Sigma$ be zero and still
have bound states. This can be seen from eqs.~(\ref{deriv_gf}) and (\ref{gf_inf_pspin}) since
a large negative $A(r)$ at large $r$ will become positive when $r<r_c$ (where $r_c$ is such that
$\Delta(r_c)=E+mc^2$) so it is clear that there exist values of $r<r_c$ such that
$A(r)f^2_\kappa> B(r)g^2_\kappa=(E-mc^2) g^2_\kappa$,
thus allowing $g_\kappa f_{\tilde\kappa}$ to change sign and fulfilling the necessary conditions to have
bound state solutions.

In refs. \cite{Chen_Meng,harm_osc_prc_2004} was shown that indeed one can have bound solutions
for a harmonic oscillator $\Delta$ potential when there is pseudospin symmetry. In this section we have proved quite generally
that this is possible for \textit{any} confining scalar and vector central potentials,
i.e., that go to $+\infty$ when $r\to\infty$.

\section{Bound solutions for anti-fermions}

If one considers the spin and pseudospin symmetries for anti-fermions, one has to consider the charge-conjugated
Dirac equation with central confining scalar and vector potentials. The effect of charge conjugation is basically
to perform the transformations $E\to -E$, $\Sigma\to -\Delta$ and $\Delta\to -\Sigma$, \cite{ronai_wscc,Zhou}.
One gets
\begin{subequations}
\label{eq_radiais_cc}
\begin{align}
\frac{d \bar g_\kappa}{dr} + \kappa\, \frac{\bar g_\kappa}{r} & =  \bar A(r)\,\bar f_{\tilde\kappa}\\
\frac{d \bar f_{\tilde\kappa}}{dr} - \kappa\, \frac{\bar f_{\tilde\kappa}}{r} & =  -\bar B(r)\,
\bar g_\kappa \ ,
\end{align}
\end{subequations}
where
\begin{subequations}
\label{def_A_B_cc}
\begin{align}
\bar A(r) &= \frac1{\hbar c}\,[-E+mc^2+\Sigma(r)] \\
\bar B(r) &= \frac1{\hbar c}\,[-E-mc^2+\Delta(r)]\
\end{align}
\end{subequations}
and $\bar g_\kappa$ $\bar f_{\tilde\kappa}$ are the radial function for the bound anti-fermions.
One can see immediately that the role of the $\Sigma$ and $\Delta$ potentials is reversed relative
to positive energy case, so that now $\Delta$ is the binding potential and $\Sigma$ is the ``mass"
potential. In our work concerning the harmonic oscillator potential in the $1+1$ Dirac
equation \cite{prc_73_054309}, but whose conclusions are valid in $3+1$ dimensions,
we have shown that one can have bound solutions in spin and pseudospin conditions for
anti-fermion for negative oscillator harmonic potentials, i.e., which go to $-\infty$
when $r\to\infty$ and that the negative energy solutions are such that $E<-mc^2$.
Therefore, considering now the case of a general confining potential
going to $-\infty$, we see that all the reasoning in the previous sections can be repeated
with $\Delta$ and $\Sigma$ being finite potentials at the origin and going to
$-\infty$ at large distances. One has $-E+mc^2>0$, $-E-mc^2>0$, and thus $\bar A(r)$ and
$\bar B(r)$ in (\ref{def_A_B_cc}) are positive at the nodes of $\bar g_\kappa$ $\bar f_{\tilde\kappa}$ and
when they are constants in pseudospin or spin symmetry conditions their asymptotic values
are such that their product is negative. Therefore, the node structure of
$\bar g_\kappa$ $\bar f_{\tilde\kappa}$ is the same as $g_\kappa$ $f_{\tilde\kappa}$
given by (\ref{nf_spin}) and (\ref{nf_pspin}), but now in reverse conditions, i.e.,
in pseudospin and spin symmetry conditions respectively. This is consistent with the finding that
bound states of anti-nucleons obtained by charge conjugation for nuclear mean field potentials
have the same radial node structure than their positive energy counterparts \cite{ronai_wscc,Zhou}.

\section{Conclusions}

In this paper we have derived the node structure of the radial functions of the upper and lower
components of a spinor which is a bound solution of the Dirac equation with scalar $S$ and
vector $V$ potentials such that $V=S$ (spin symmetry) and $V=-S$ (pseudospin symmetry), when
those potentials are finite at the origin and go to $+\infty$ when $r\to\infty$,
independently of their shape.
It was shown that the node structure when pseudospin symmetry exists is different from the usual
node structure in systems of scalar and vector potentials with are finite at the origin and
go to zero at infinity. However, in spin symmetry conditions, we found that the node structure
for confining potentials is the same as for potentials which go to zero at infinity.

We have also shown that, contrary to what happens with potentials which go to zero at large distances,
confining potentials allow to have bound states in pseudospin symmetry conditions for positive energy
states and in spin symmetry conditions for negative energy states.

We believe that the proof presented here of the existence of positive energy bound states for any radial confining potential
in the limit of pseudospin symmetry is quite relevant, since it uncovers a whole class of potentials which allow to realize
exactly this symmetry in nature.
While it can realized only approximately in physical systems governed by interactions going to zero at large distances,
as in relativistic mean-field theories for nuclei or in atoms, one can find other systems where either harmonic oscillator or linear confining
potentials are crucial for explaining their properties, as, for example, the Cornell potentials in particle physics.
Very recently, solutions of the Dirac equations with scalar, vector and tensor Cornell radial potentials in the spin and pseudospin symmetry limit have been obtained, and it was shown that the solution in each symmetry limit case can be related by a chiral transformation \cite{Benito}, as was already been shown for harmonic oscillator potentials \cite{prc_73_054309}. In that work it is explicitly shown that the radial Dirac equation for this problem can be  mapped into a Schrodinger-like equation with a harmonic oscillator plus a Cornell potential whose solution is presented. This potential is known in particle physics as the Killingbeck potential \cite{Killingbeck} and was also considered recently in the context of the Dirac equation and its solutions in the limit of the spin and pseudospin symmetries were obtained \cite{Hamzavi_1,Hamzavi_2}. Thus we believe that the results presented here can be applied to a wide range of physical systems.

\begin{acknowledgments}
We acknowledge financial support from CNPq and QREN/FEDER,
the Programme COMPETE, under Project No.~PTDC/FIS/113292/2009.
P. Alberto would like also to thank the Universidade Estadual
Paulista, Guaratinguet\'a campus, for supporting his stays in its
Physics and Chemistry Department.
M.~Malheiro acknowledges CNPq, Capes/FCT Brazilian-Portuguese scientific agreement,
and FAPESP for the financial support.
\end{acknowledgments}

\end{document}